\documentclass[pra,twocolumn,showpacs]{revtex4}
\usepackage{bm,graphicx,amsmath}

\begin{document}

\title{Coupled two-component atomic gas in an optical lattice}
\author{Jonas Larson and Jani-Petri Martikainen}
\affiliation{NORDITA, 106 91 Stockholm, Sweden}
\date{\today}

\begin{abstract} 
We present an {\it ab initio} study of the ground state of an ideal coupled two-component gas of ultracold atoms in a one dimensional optical lattice, either bosons or fermions. Due to the internal two-level structure of the atoms, the Brillouin zone is twice as large as imposed by the periodicity of the lattice potential. This is reflected in the Bloch dispersion curves, where the energy bands regularly possess several local minima. As a consequence, when the system parameters are tuned across a resonance condition, a non-zero temperature phase transition occurs which arises from an interplay between internal and kinetic atomic energies. For fermions, this phase transition is of topological character since the structure of the Fermi surface is changed across the critical value. It is shown that these phenomena are also expected to occur for two and three dimensional optical lattices. 
\end{abstract}

 \maketitle

\section{Introduction} 
The coupled dynamics between an ultracold atomic gas and an optical lattice has gained enormous attention during the last decade. Originally, D. Jaksch {\it et al.} proposed that the {\it Mott-superfluid} quantum phase transition (PT) can be realized using cold atoms dipole coupled to a standing wave laser field \cite{jaksch}. A mere four years later, this PT was observed in the seminal experiment by I. Bloch and coworkers \cite{bsf}. One among the many reasons for the great interest in these systems is the possibility to experimentally realize, and therefore verify, theoretical models developed in the field of condensed matter physics, see the review \cite{maciek}. Several achievements in this field have been accomplished both theoretically and experimentally, and to only mention a few; disordered systems and matter localization \cite{dis}, frustrated systems \cite{frust}, the strong atom-atom interaction regime and the Tonks gas \cite{tonks}, cold atom quantum Hall counterp
 arts \cite{hall} and quantum information processing \cite{qi}. 

Almost exclusively, effective dispersive models have been considered, where the internal level structure of the atoms can be discarded and one is left with a single atomic level. Exceptions are works on spin-dependent atoms, which have shown to acquire new phases compared to the spinless case \cite{spin1}. Another, related system is mixtures of different atomic species, again containing novel physics \cite{mix}. However, these models do not analyze direct coupling between the internal atomic states, as will be the topic of the present paper. Furthermore, atom-atom scattering plays a crucial role in the dynamics of all above references. For example, the Mott-superfluid PT in the {\it Bose-Hubbard model} arises from a competition between the atomic kinetic energy (hopping between neighboring sites) and the onsite scattering interaction \cite{sachdev}. Here we will show that by taking into account for dipole induced transitions between coupled atomic states, PTs can occur even f
 or ideal gases lacking atom-atom interaction. This originates from the competition between the atomic kinetic and the internal energies, in comparison with kinetic and scattering energies.

Consequently, we do not restrict the character of the atom-field interaction to be dispersive. In fact, we allow for a vanishing atom-field detuning and in particular study the behavior for both positive and negative detunings. Coupling between internal levels of atoms in optical lattices is rather unexplored. The first analysis seem to date back to the works by Krutitsky {\it et. al.}, where they considered effective coupled two or three level models derived from Raman interactions \cite{multicomp1,multicomp2}. They found that the hopping term in the corresponding Bose-Hubbard model may be tuned to be either positive or negative, leading to new phases. An internal coupled model was as well studied by Garc\`ia-Ripoll {\it et. al.} in Ref.~\cite{GR}. Contrary to \cite{multicomp1,multicomp2}, we consider here a direct coupling between two atomic levels, which is actually what one gets in the standard optical lattice model by decreasing the amplitude of the detuning. Another crucial difference between the current contribution and the ones of Ref.~\cite{multicomp1,multicomp2} is that our model goes beyond any {\it tight binding approximation}, in which hoppinig beyond nearest neighbours is allowed. The detuning in our model may be seen as having similar role as the parameter $\theta$ (relative phase between the two Raman lasers) in \cite{multicomp1}. An important point to note is that in Ref.~\cite{GR}, on the other hand, the coupling between internal levels is not originating from a non-zero detuning (as in the current paper) or from a non-zero angle $\theta$ (as in \cite{multicomp1,multicomp2}), but from atom-atom scattering. We explicitly show that the two-level structure of our model renders dispersion curves having local minima. It is known that such phenomena can give rise to topological PTs \cite{top,top2,top3}, which indeed are found in our model as well. A topological PT (for fermions; a topological change in the Fermi surface) does take place when the detuning changes sign. The same kind of PT indeed occurs also for bosons. The non-zero temperature situation is also considered and the topological PT is found to be stable against temperature fluctuations. 

The outline of this paper is as follows. We first discuss the single particle Hamiltonian in Sec.~\ref{sec2}, and point out some of the symmetries associated with it. One of these symmetry operators indicates that the Brillouin zone extends over twice the size expected from the periodicity of the optical lattice, a fact that is clarified even further in the proceeding Sec.~\ref{sec3}. In Sec.~\ref{sec3} we begin by analyzing the energy spectrum and the two lowest band Wannier functions. Using this knowledge we demonstrate the presence of a PT, both for fermions and bosons at zero and non-zero temperatures. We also discuss the effect of weak atom-atom interaction. A discussion of possible extensions is left for the conclusions given in \ref{sec4}. Last, in the appendix \ref{app} we also present an effective Raman coupled model that would provide qualitatively the same results, but with the benefit of using two meta-stable atomic states. 

\section{Single particle Hamiltonian}\label{sec2}
The Hamiltonian for a single two-level atom, whose internal states dipole couple via a standing wave laser field, reads
\begin{equation}\label{ham1}
\hat{H}=\frac{\hat{\tilde{p}}^2}{2m}+\frac{\hbar\tilde{\Delta}}{2}\hat{\sigma}_z+2\hbar\tilde{g}\cos(k\hat{\tilde{x}})\hat{\sigma}_x.
\end{equation}
Here, $\hat{\tilde{p}}$ and $\hat{\tilde{x}}$ are atomic center-of-mass momentum and position, $m$ its mass, $\tilde{\Delta}$ the atom-field detuning, $\tilde{g}$ the effective atom-field coupling and $\tilde{k}$ the field wave number. The internal states of the atom are labeled $|\pm\rangle$ and the Pauli matrices operate as $\hat{\sigma}_z|\pm\rangle=\pm|\pm\rangle$ and $\hat{\sigma}_x|\pm\rangle=|\mp\rangle$. Before proceeding we introduce dimensionless variables through the characteristic length $k^{-1}$ and energy $E_r=\hbar^2k^2/2m$;
\begin{equation}
\begin{array}{lllll}
\hat{x}=k\hat{\tilde{x}}, & & \displaystyle{\Delta=\frac{\hbar\tilde{\Delta}}{E_r}}, & & \displaystyle{g=\frac{\hbar\tilde{g}}{E_r}}.
\end{array}
\end{equation}
In the $|+\rangle=\left[\begin{array}{c}1\\ 0\end{array}\right]$ and $|-\rangle=\left[\begin{array}{c}0\\ 1\end{array}\right]$ nomenclature, Eq.~(\ref{ham1}) becomes in scaled variables
\begin{equation}\label{ham2}
\hat{H}=-\frac{\partial^2}{\partial x^2}+\left[\begin{array}{cc}
\displaystyle{\frac{\Delta}{2}} & 2g\cos(\hat{x}) \\ 
2g\cos(\hat{x}) & -\displaystyle{\frac{\Delta}{2}}\end{array}\right],
\end{equation} 
which serves as our model Hamiltonian. We note that the external field couples the {\it bare states} $|\pm\rangle$ and simultaneously shift the momentum by either $\pm1$. For $\Delta=0$, the unitary operator $\hat{U}=\frac{1}{\sqrt{2}}\left(\hat{\sigma}_x+\hat{\sigma}_z\right)$ decouples the internal levels, and one obtains two Mathieu equations \cite{mat} with {\it diabatic potentials} $V_\pm^d(x)=\pm 2g\cos(x)$ \cite{dong}. The internal atomic states of the decoupled equations are
\begin{equation}\label{bas2}
\begin{array}{l}
\displaystyle{|1\rangle=\hat{U}|+\rangle=\frac{1}{\sqrt{2}}(|+\rangle+|-\rangle)},\\ \\
\displaystyle{|2\rangle=\hat{U}|-\rangle=\frac{1}{\sqrt{2}}(|+\rangle-|-\rangle)}.
\end{array}
\end{equation}

As a periodic operator, the eigenstates and eigenvalues of Hamiltonian (\ref{ham2}) are characterized by two quantum numbers, one discrete band index $\nu$ and a continuous quasi momentum $q$
\begin{equation}\label{blochb}
\hat{H}|\psi_\nu(q)\rangle=E^\nu(q)|\psi_\nu(q)\rangle,\hspace{1cm}\nu=1,\,2,\,3,...\,.
\end{equation}
The range of $q$ assigns the first Brillouin zone and is determined by the symmetries of the problem. For any parameters, the operator $\hat{T}=\mathrm{e}^{i\lambda\hat{p}}$ commutes with the Hamiltonian, where $\lambda=2\pi$ is the field wavelength in scaled units. Not as evident is that also the operator
\begin{equation}
\hat{I}=\hat{\sigma}_z\mathrm{e}^{i\frac{\lambda}{2}\hat{p}}
\end{equation}
is a symmetry of the Hamiltonian \cite{hamsym,jonaseff}. Thus, there is a natural $\pi$ periodic structure of the Hamiltonian. This signals that the first Brillouin zone extends beyond $-1/2<q\leq1/2$ as implied by the $2\pi$ periodicity of the $\hat{T}$ operator, and implicitly we have
\begin{equation}
q\in(-1,1].
\end{equation}
However, for $\Delta=0$ the spectrum is doubly degenerate and in this limit it might be more convenient to define the Brillouin zone within $-1/2<q\leq1/2$. One may note that $\hat{T}=\hat{I}^2$. The physical background of this additional symmetry operator $\hat{I}$ emerges from the fact that absorption or emission of a single photon flips the internal states, $|\pm\rangle\rightarrow|\mp\rangle$, while only for a two-photon process is the internal state unchanged. 

\section{Structure of the ground state}\label{sec3}
\subsection{Model characteristics}\label{ssec3a}
To find the ground state of $N$ non-interacting two-level atoms we need to diagonalize (\ref{ham2}); both within the internal and the motional degrees of freedom. We first note that for a given quasi momentum $q$, the Hamiltonian can be written on the block form $H=H_\varphi\otimes H_\phi$, and the states coupled by the sub-Hamiltonians read
\begin{equation}\label{basis}
\begin{array}{l}
|\varphi_\eta(q)\rangle=\left\{\begin{array}{l}
|q+\eta\rangle|-\rangle\hspace{1cm}\eta\,\,\mathrm{even}\\
|q+\eta\rangle|+\rangle\hspace{1cm}\eta\,\,\mathrm{odd}\end{array}\right.\\ \\
|\phi_\eta(q)\rangle=\left\{\begin{array}{l}
|q+\eta\rangle|+\rangle\hspace{1cm}\eta\,\,\mathrm{even}\\
|q+\eta\rangle|-\rangle\hspace{1cm}\eta\,\,\mathrm{odd},\end{array}\right.
\end{array}
\end{equation}
where the first ket is the momentum eigenstate $\hat{p}|q+\eta\rangle=(q+\eta)|q+\eta\rangle$ and $\eta$ is an integer corresponding to the momentum shift governed by absorption and emittance of photons.  This decoupling of the dynamics is of great importance as a given number of atoms residing in each subset is constant when the parameters are varied. The states (\ref{basis}) are eigenstates of the Hamiltonian in the absence of matter-field coupling $g=0$, with {\it bare energies}
\begin{equation}
\varepsilon^\mu=(q+\eta)^2\pm(-1)^\mu\frac{\Delta}{2},
\end{equation}
where the $\pm$-sign is different for the two sets (\ref{basis}). In this trivial situation, the detuning $\Delta$ simply shifts the parabolic dispersions. The general eigenstate, {\it Bloch state}, were introduced in Eq.~(\ref{blochb}). The eigenvalue $E^\nu(q)$ is the $\nu$'th energy band/dispersion curve, given for $q\in(-1,1]$. The eigenstate in a position representation can be written using its {\it constituent Bloch states} as
\begin{equation}
\begin{array}{lll}
\psi_{\nu,q}(x)=\langle x|\psi_\nu(q)\rangle & = & \psi_{\nu,q}^+(x)|+\rangle+\psi_{\nu,q}^-(x)|-\rangle\\ \\
& = & \psi_{\nu,q}^1(x)|1\rangle+\psi_{\nu,q}^2(x)|2\rangle.
\end{array}
\end{equation}
Note that the above specifies just two possible constituent Bloch states and any other internal basis would define two new ones. In particular, $\psi_{\nu,q}^{1,2}(x)$ and $\psi_{\nu,q}^\pm(x)$ are related via the operator $\hat{U}=\frac{1}{\sqrt{2}}\left(\hat{\sigma}_x+\hat{\sigma}_z\right)$ of Eq.~(\ref{bas2}). In terms of the the dispersions, the Hamiltonian is given by
\begin{equation}
\hat{H}=\sum_{\nu=1}^\infty\sum_{q\in(-1,1]}\big[E^\nu(q)-\mu\big]\hat{n}_q^\nu,
\end{equation}
where $\hat{n}_q^\nu$ is the operator giving the number of atoms with energy $E^\nu(q)$ and here we have introduced a chemical potential $\mu$. Note that for a given atom number $N$ one has $\sum_{\nu=1}^\infty\sum_{q\in(-1,1]}\langle\hat{n}_q^\nu\rangle=N$, where $\langle\hat{n}_q^\nu\rangle$ is the expectation value for the particular state of interest. The Bloch states can be expressed in the {\it bare basis} (\ref{basis})
\begin{equation}\label{coef1}
\begin{array}{lll}
|\psi_\nu(q)\rangle & = & \left\{\begin{array}{l}\displaystyle{\sum_{\eta=-\infty}^\infty c_\eta^\nu(q)|\varphi_\eta(q)\rangle} \\ \\
\displaystyle{\sum_{\eta=-\infty}^\infty d_\eta^\nu(q)|\phi_\eta(q)\rangle}
\end{array}\right..
\end{array}
\end{equation}
Note that expanding $|\psi_\nu(q)\rangle$ does not mix the states $|\varphi_\eta(q)\rangle$ and $|\phi_\eta(q)\rangle$ due to the block-diagonal form of the Hamiltonian $H=H_\varphi\otimes H_\phi$ in this bare basis. We emphasized that a quasi momentum eigenstate, or Bloch state, is a linear combination of the two internal atomic states $|\pm\rangle$. The same holds for the Wannier functions
\begin{equation}\label{wann}
\begin{array}{lll}
w_\nu(x-R) & \equiv & \displaystyle{\sum_{q\in(-1,1]}\mathrm{e}^{iqR}\psi_{\nu,q}(x)}\\ \\ & = & w_\nu^+(x-R)|+\rangle+w_\nu^-(x-R)|-\rangle\\ \\
& = & w_\nu^1(x-R)|1\rangle+w_\nu^2(x-R)|2\rangle.
\end{array}
\end{equation} 
As defined above, the {\it constituent Wannier functions} $w_\nu^+(x-R)$ and $w_\nu^-(x-R)$ are normalized as $\sum_{i=\pm}\int |w_\nu^i(x-R)|^2dx=1$, and likewise for $w_\nu^1(x-R)$ and $w_\nu^2(x-R)$. Like for the Bloch states, the constituent parts depend on the internal basis. Equipped with this machinery, we now turn to analyze the complex band structure of the Hamiltonian (\ref{ham2}).

\begin{figure}[ht]
\begin{center}
\includegraphics[width=8cm]{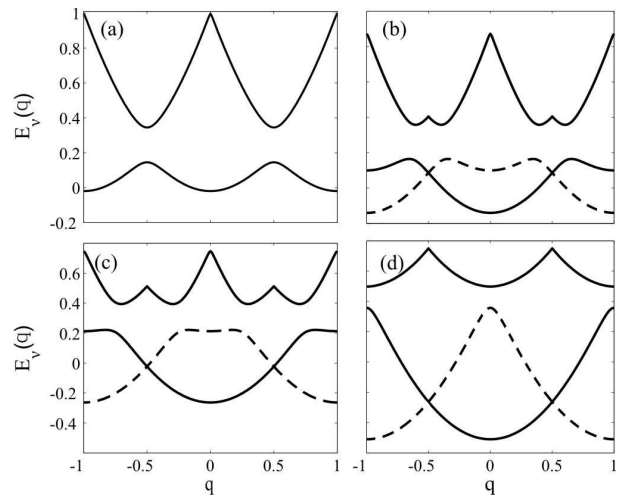}
\caption{The three lowest energy bands of Hamiltonian (\ref{ham2}), where the scaled dimensionless parameters are $g=0.1$ and $\Delta=0,\,0.25,\,0.5,\,1$ in plots (a)-(d) respectively. The two lowest bands are distinguished between the two sub-Hamiltonians $H_\varphi$ (solid line) and $H_\phi$ (dashed line).  }
\label{fig1}
\end{center}
\end{figure}

Figure \ref{fig1} displays some examples of the band structure - the dispersion curves, where the first three bands are shown. The resonance case, $\Delta=0$, is given in (a) and here the energy spectrum is doubly degenerate as pointed out in the previous section. However, for non-zero $\Delta$, the degeneracy is lifted (indicated in (b)-(d)) and as a consequence, the first Brillouin zone extends over quasi momenta $q\in(-1,1]$. The dashed line corresponds to the lowest dispersion curve of the sub-Hamiltonian $H_\phi$, while the other lowest line derives from $H_\varphi$ instead. Note that the dispersion curves have more than a single minimum in all plots. The non-monotonicity of the dispersion curves can be understood from the interplay between the different energies; kinetic energy causing the parabolic structures, the coupling energy which typically splits the degeneracy and the internal atomic energy that (loosely speaking) shifts the dispersions.  

Let us briefly discuss the dispersion curves and their limiting cases in more detail. In the regular dispersive situation one has $\Delta\gg g$ and the coupling between the internal adiabatic states can be neglected. To verify this, let us introduce the {\it adiabatic states} \cite{jonasad} as columns of the unitary operator
\begin{equation}
\hat{U}_{ad}=\left[\begin{array}{cc} \cos(\theta/2) & \sin(\theta/2) \\
\sin(\theta/2) & -\cos(\theta/2)\end{array}\right],
\end{equation}
where
\begin{equation}
\tan(\theta)=\frac{4g\cos(x)}{\Delta}.
\end{equation}    
The operator $\hat{U}_{ad}$ then diagonalizes the $2\times2$ matrix of the Hamiltonian (\ref{ham2}). However, due to its $x$-dependence will it not commute with the momentum operator $\hat{p}$. This causes non-diagonal terms in the transformed Hamiltonian \cite{jonasad},
\begin{equation}
\begin{array}{lll}
\tilde{H} & = & \displaystyle{\hat{U}\hat{H}\hat{U}^{-1}=-\frac{\partial^2}{\partial x^2}+V_{cent}(\hat{x})}\\ \\
& & +\left[\begin{array}{cc}V_+^{ad}(\hat{x}) & \Omega(\hat{x},\hat{p})\\ 
\Omega^*(\hat{x},\hat{p}) & V_-^{ad}(\hat{x})\end{array}\right].
\end{array}
\end{equation}
Here, $V_{cent}(x)$ is a {\it centrifugal} term turning up on the diagonal \cite{cent}, $\Omega(x,p)$ is the {\it non-adiabatic coupling} \cite{jonasad} and specifically
\begin{equation}
V_\pm^{ad}(x)=\pm\sqrt{\left(\frac{\Delta}{2}\right)^2+4g^2\cos^2(x)}
\end{equation}
are the {\it adiabatic potentials}. It follows that the centrifugal and adiabatic correction terms are small in the $\Delta\gg g$ regime \cite{jonasad}, giving the adiabatic potentials
\begin{equation}
V_\pm^{ad}(x)\approx\pm\frac{\Delta}{2}\pm\frac{4g^2\cos^2(x)}{\Delta}.
\end{equation}
Thus, we derive the regular situation most commonly considered in the literature and we especially note that the first Brillouin zone extends over $q\in(-1,1]$. Choosing $\Delta>0$ and considering the weak coupling limit $4g^2/\Delta\rightarrow0$, the states $|q\rangle|-\rangle$ with $q\in(-1,1]$ have eigenvalues
\begin{equation}
E^0(q)=\varepsilon^0=q^2-\frac{\Delta}{2}.
\end{equation}
For the states $|q\pm1\rangle|-\rangle$ on the other hand, we have the dispersions
\begin{equation}
E^0(q)=(q\pm1)^2-\frac{\Delta}{2}.
\end{equation}
However, as is noticeable from Eq.~(\ref{basis}), the states $|q\rangle|-\rangle$ and $|q\pm1\rangle|-\rangle$ are not coupled and when one restricts the analysis to just one of these sets and one consequently regains a regular spectrum, by which we mean dispersion curves that posses only a single minimum within one Brillouin length, $dE^1(q)/dq\geq0$ for $0\leq q\leq1$. More interesting is the intermediate regime, neither adiabatic nor {\it diabatic} ($\Delta=0$), where the above states may indeed coexist if we assume that both of the states $|q\rangle|\pm\rangle$ can be present. Thus, when the coupling between $|\pm\rangle$ states cannot be neglected, one must take into account all the states and the lowest energy band contains minima at $q=\pm1$ as well as for $q=0$. 

As pointed out, here we allow for atomic states $|+\rangle$ and $|-\rangle$ to have the same momentum. It is understood that the same arguments hold for $\Delta<0$, making the replacement $|+\rangle\leftrightarrow|-\rangle$. We should mention that even if we only consider atoms within one subset (\ref{basis}), the dispersion curves will contain several local minima \cite{jonaseff,jonasbloch}, and the results presented are valid also in such cases. In the diabatic limit, $\Delta=0$, we saw that we can separate the dynamics into two uncoupled problems with {\it diabatic potentials} $V_\pm^{d}(x)=\pm2g\cos(x)$. The spectrum is then doubly degenerate and the first Brillouin zone is therefore most properly defined within $q\in(-1/2,1/2]$. Noteworthy is the fact that the diabatic potentials have minima either for $x_m^{d^-}=2n\pi$ or $x_m^{d^+}=(2n+1)\pi$ for integer $n$, while the minima for the adiabatic potentials are either $x_m^{a^-}=n\pi$ or $x_m^{a^+}=\left(n+\frac{1}{2}\right)\pi$. By the $\pm$-sign we indicate the corresponding diabatic or adiabatic potential.

\begin{figure}[ht]
\begin{center}
\includegraphics[width=8cm]{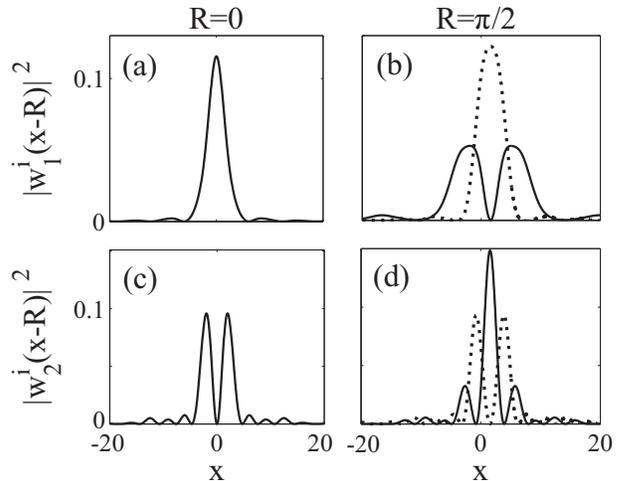}
\caption{The constituent squared amplitudes of the Wannier functions of the first, (a) and (b), and second, (c) and (d), Bloch band. The left plots display the Wannier functions for $R=0$, while the in the right plots $R=\pi/2$, corresponding to the minima of $V_-^{ad,d}(x)$ and $V_+^{ad}(x)$ respectively. Solid lines depict $|w_1^+(x-R)|^2$ and dashed line $|w_1^-(x-R)|^2$. The other dimensionless parameters are $g=0.1$ and $\Delta=0.25$. }
\label{fig2}
\end{center}
\end{figure}

The coupled two-level character of the system gives rise to rather peculiar constituent Wannier functions $w_\nu^\pm(x-R)$ or $w_\nu^{1,2}(x-R)$. In general, $R$ is chosen such that the potential is minimal at $x=R$. However, in the present model $R$ is not {\it a priori} given since the potential has a complex two-level structure. From the symmetry of the Hamiltonian it follows that $w_\nu^+(x-R)\leftrightarrow w_\nu^-(x-R)$ for $\Delta\leftrightarrow-\Delta$, while $|w_\nu^{1,2}(x-R)|$ are invariant under such sign change. In Fig.~\ref{fig2} (a)-(d) we show examples of $|w_1^\pm(x-R)|^2$ (a) and (b) and $|w_2^\pm(x-R)|^2$ (c) and (d), for $R=0$ or $R=\pi/2$. The dashed line shows $|w_1^-(x-R)|^2$ and the solid $|w_1^+(x-R)|^2$. Note that for $R=0$, the two constituent Wannier functions are identical, which, however, does not hold for $R=\pi/2$. For $R=0$, the constituent Wannier functions resemble typical ones obtained in one component systems \cite{mermin}. This is not tru
 e for $R=\pi/2$ (corresponding to the minima of $V_+^{ad}(x)$), where only $|w_\nu^-(x-R)|^2$ shows the regular shape and $|w_1^+(x-R)|^2$ looks as if $R$ coincide with a maximum of the potential. Note that the result of the figure presents an intermediate regime where neither the adiabatic nor the diabatic approximations can be considered; $g=0.1$ and $\Delta=0.3$. 

\begin{figure}[ht]
\begin{center}
\includegraphics[width=7cm]{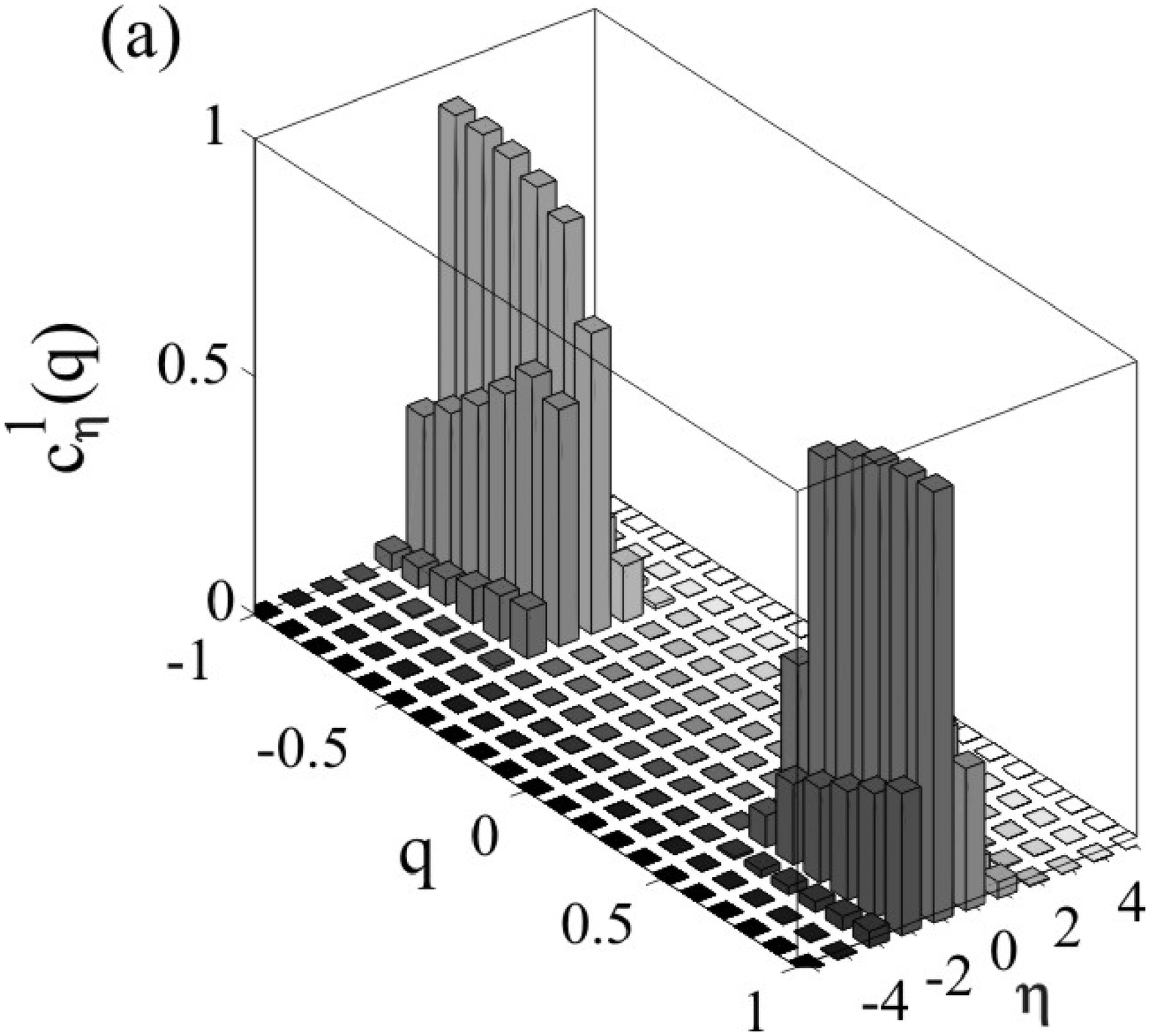}
\includegraphics[width=7cm]{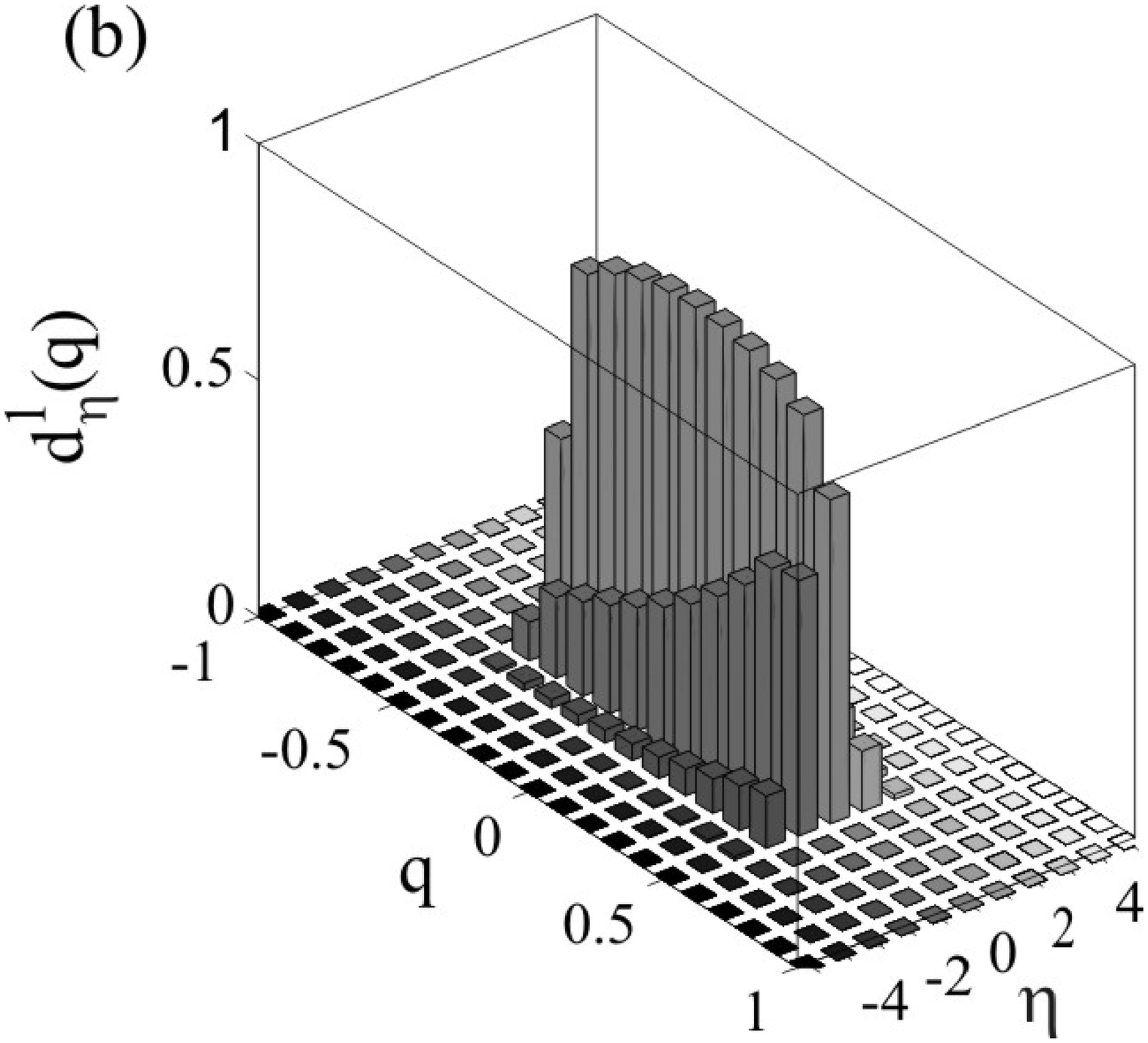}
\caption{Coefficients $c_\eta^1(q)$ (a) and $d_\eta^1(q)$ (b) defined in Eq.~(\ref{coef1}). The parameters are as in Fig.~\ref{fig1} (b), but with a larger coupling $g=0.5$. }
\label{fig3}
\end{center}
\end{figure}

\subsection{Zero temperature many-body ground state}\label{ssec3b}
We will focus on the lowest band, $\nu=1$, thus assuming a {\it filling factor} $\xi=\frac{N}{K}<1$, where $N$ is the total number of atoms and $K$ number of sites. In the numerics, $K$ will be taken large enough (typically $K>100$) to assure small boundary effects. The $N$ many-body ground states for fermions and bosons, at zero temperature and for filling factors $\xi<1$, are 
\begin{equation}\label{fermisea}
|\Psi\rangle_{F}=\prod_{q\in \mathcal{Q}}\hat{f}_q^\dagger|0\rangle=\prod_{q\in\mathcal{Q}}|n_q\rangle,
\end{equation}
\begin{equation}\label{bosstate}
|\Psi\rangle_B=|n_0,n_{+1}\rangle,\hspace{1cm}n_0+n_{+1}=N
\end{equation}
respectively and $\mathcal{Q}$ contains those values of $q$ which are inside the Fermi
sea, $\hat{f}_q^\dagger$ is the Fermi creation operator of mode $q$ of the lowest Bloch band, $|0\rangle$ is the vacuum and  $n_q$ ($=\langle\hat{n}_q\rangle$) again characterizes the number of atoms in quasi momentum mode $q$; $\sum_{q\in\mathcal{Q}}n_q=N$. Note that for bosons, the ground state is degenerate and $n_0$ and $n_{+1}$ may pertain any positive values such that the total number gives $N$ (we have not included $q=-1$ as it lies outside the Brillouin zone), and in fact any linear combination of these degenerate states is adequate. Further, the state (\ref{bosstate}) is valid for any filling factors $\xi$. In fact, the same holds in general for any non-integer filling factor $\xi$. As 
states (\ref{fermisea}) and (\ref{bosstate}) are prodcut states, 
there exist no correlation between the atoms. Nonetheless, entanglement between atomic motion and internal atomic states occurs for each atom (provided $\Delta,\,g\neq0$). This  is clear since
\begin{equation}
\langle x|\psi_\nu(q)\rangle\neq\chi(x)\left(a|+\rangle+b|-\rangle\right),
\end{equation}
for some normalized function $\chi(x)$ and $|a|^2+|b|^2=1$. The expansion of the Bloch eigenstates in terms of bare states is given in Eq.~(\ref{coef1}). The corresponding coefficients are displayed in Fig.~\ref{fig3}, for the lowest Bloch band with parameters $g=0.5$ and $\Delta=0.25$. For $|q|<1/2$, the $c_\mu^1(q)$ are all zero, while $d_\mu^1(q)=0$ for $1/2<|q|<1$. A result deriving from the block structure of the Hamiltonian (\ref{basis}). The $c_\mu^1(q)$ coefficients are dominated by the ones with either $\mu=1$ or $\mu=-1$, and $\mu=0$ is the most prominent coefficient of the $d_\mu^1(q)$. Note that the probability for the atom to be found in the state $|-\rangle$ for a given eigenstate $\psi_\nu(q)$ is
\begin{equation}
P(-;\psi_\nu(q))=\sum_{\eta\,\,\mathrm{even}}|c_\eta^\nu(q)|^2+\sum_{\eta\,\,\mathrm{odd}}|d_\eta^\nu(q)|^2
\end{equation}
and $P(+;\psi_\nu(q))=1-P(-;\psi_\nu(q))$. 

\begin{figure}[ht]
\begin{center}
\includegraphics[width=8cm]{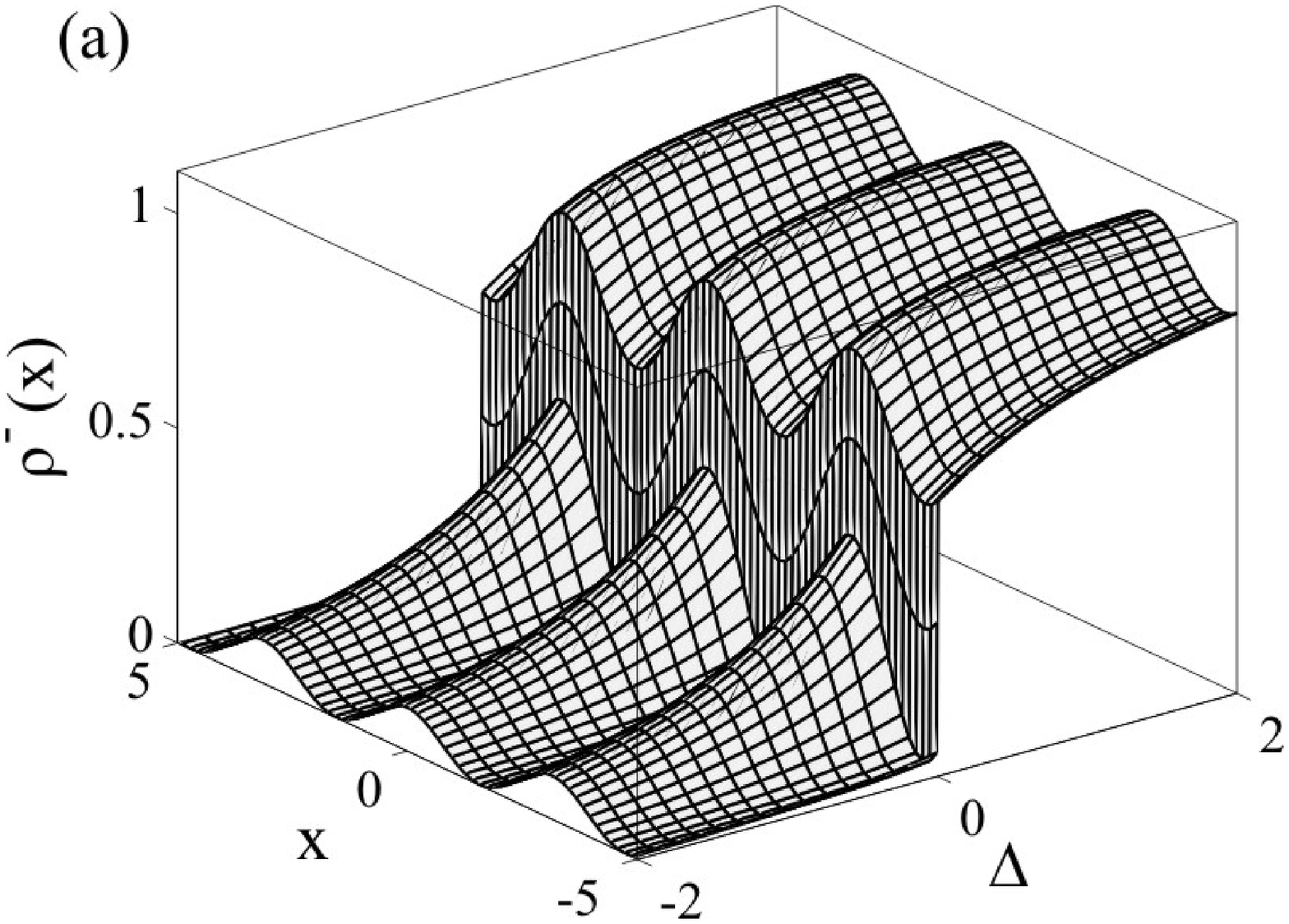}
\includegraphics[width=7cm]{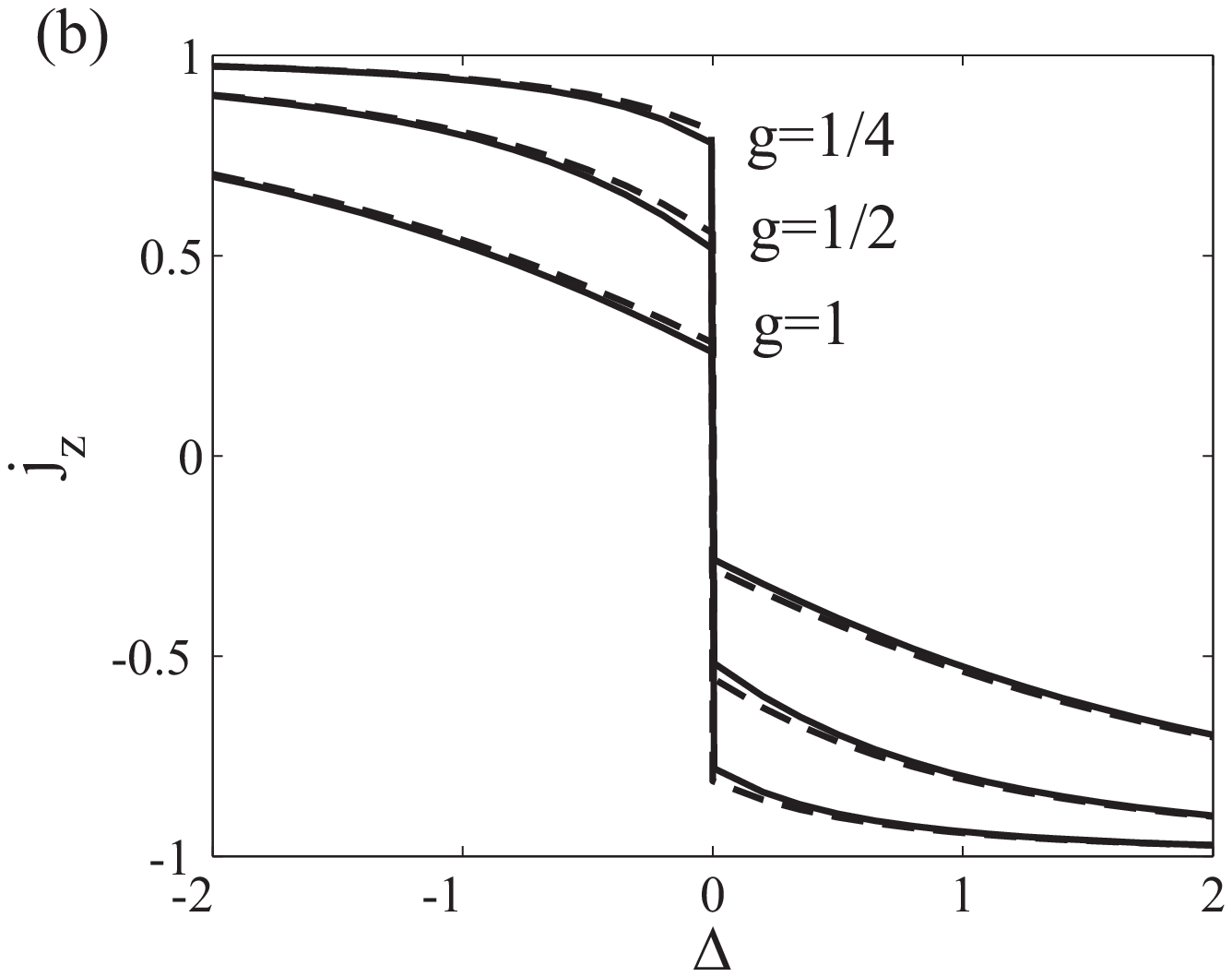}
\caption{The upper plot (a) displays the Fermi particle density $\rho^-(x)$ (\ref{density}) for the internal state $|-\rangle$ at half filling of the lowest Bloch band. The lower plot (b) shows three examples of the collective atomic inversion $j_z$ (\ref{atinv}) as function of $\Delta$ for bosons (dashed lines) and for fermions (solid lines). In (a) $g=1$.   }
\label{fig4}
\end{center}
\end{figure}   

We define the internal density per particle as
\begin{equation}\label{density}
\rho^\pm(x)=\frac{1}{N}\sum_\nu\sum_{q\in\mathcal{Q}}|\psi_{\nu,q}^\pm(x)|^2,
\end{equation}
where the second sum runs over occupied momentum states, and $N$ is the number of atoms. The density $\rho^-(x)$ for fermions is shown in Fig.~\ref{fig4} (a) as function of the detuning $\Delta$ where the lowest Bloch band is half filled, $\xi=1/2$. For other filling factors one regains very similar plots. By substituting $\Delta\leftrightarrow-\Delta$, the $\rho^+(x)$ is identical to $\rho^-(x)$. Noticeable is the discontinuity at $\Delta=0$. Thus, around resonance, $\Delta=0$, the population of $|+\rangle$ and $|-\rangle$ atoms may fluctuate and in particular a first order PT in the {\it collective atomic inversion} (per particle)
\begin{equation}\label{atinv}
j_z\equiv\frac{1}{N}\sum_\nu\sum_{\{n_q\}}\left[P(+;\psi_\nu(q))-P(-;\psi_\nu(q))\right]
\end{equation}
is expected. Note that the above inversion (\ref{atinv}) may be derived from the expectation value of the collective inversion $\hat{J}_z$
\begin{equation}
j_z=\frac{\langle\hat{J}_z\rangle}{N}\equiv\frac{1}{N}\sum_{i=1}^N\langle\hat{\sigma}_z^{(i)}\rangle,
\end{equation}
where $\hat{\sigma}_z^{(i)}$ is the $i$'th atoms $z$ Pauli matrix measuring the $i$'th particle inversion. In the {\it Dicke model} of $N$ two-level atoms interacting with a quantized field mode, $j_z$ is often considered as an order parameter having a discontinuous first order derivative at the critical atom-field coupling which defines the {\it normal-superradient quantum PT} \cite{jonasdicke}. The inversion $j_z$ as function of $\Delta$, presented in Fig.~\ref{fig4} (b), illustrates the discontinuity incorporated in our model. 
Note that this PT is different from the one of the Dicke model, where the PT originates from competition between field and interaction energies and not between kinetic and internal atomic energies as in this paper. In fact, the Dicke model lacks any sort of periodic lattice potential. 

We have verified that the PT obtained in our model also occurs when one restricts the dynamics to one of the subsets (\ref{basis}). Assuming a single particle in either of the subsets (\ref{basis}), one has $E_1(\pm1)=E_1(0)$ for $\Delta=0$, while $E_1(\pm1)\neq E_1(0)$ for $\Delta\neq0$ (clear for example from the dashed curve if Fig.~\ref{fig1}). Note that this is not true if we include both subsets where $E_1(\pm1)=E_1(0)$ is always true (here $E_1(q)$ is the lowest Bloch band of either $H_\phi$ or $H_\varphi$). Thus, the single particle ground state energy is either $E_1(\pm1)$ or $E_1(0)$ depending on the sign of $\Delta$, and as one tunes $\Delta$ across resonance the atom will absorb/emit one photon and $W\rightarrow-W$. 

The same argument holds also for many atoms where each one of them swaps internal states, $|\pm\rangle\leftrightarrow|\mp\rangle$, while passing through $\Delta=0$. For a fermionic system, the PT is of topological character, since the structure of the Fermi surface changes across the critical point due to the fact that each atom emits/absorbs a momentum kick. The mechanism behind the PT is the interplay between internal and external atomic energies. The collective inversion, related to internal structure, reveals the PT, but its topological character is, however, only manifested in the external degrees of freedom. Even for bosons can the PT be termed topological becuase the nature of the atomic phase space distribtion is altered while crossing the critical point. Topological PTs, or Lifshitz transition, have been studied comprehensively in quantum Hall and superconducting systems \cite{top,top2}. However, a recent paper considered a topological PT of ultracold fermionic atoms in an anisotropic three dimensional optical lattice \cite{top3}.  

The non-linearity, imposed by including atom-atom interaction, is known to cause changes such as loops in the band structure around the avoided crossings of the dispersions \cite{pethick}. This effect occurs for rather strong couplings, or more precisely when the effective scattering amplitude becomes comparable or larger than the lattice depth \cite{pethick}. For reasonable temperatures is the Fermi surface well below the first band gap where the greatest effects of the non-linearity is expected to manifest itself. Consequently, the topological PT is supposed to be insensitive for such modifications, since the general structure of the spectrum remains and thus the topological change in the Fermi sea would still take place as the detuning is tuned across resonance. Nonetheless, one may expect that even for weak interactions, the interaction induced splitting of the degeneracy of the ground state might cause a re-formation of the ground state into a correlated state which is fundamentally different from the simple product state. Thus, the ground state is no longer a simple product state of Fock states as given in Eqs.~(\ref{fermisea}) and (\ref{bosstate}). On the other hand, the inversion is invariant for such re-configuration of the ground state because $\langle \psi_\nu(q)|\hat{\sigma}_z|\psi_\nu(q)\rangle=\langle \psi_\nu(-q)|\hat{\sigma}_z|\psi_\nu(-q)\rangle$, $q\in(-1,1]$.

One limitation of our model is spontaneous emission of the excited atomic state. This can be circumvented by considering instead a Raman coupled $\Lambda$-atom setup, in which the excited state is adiabatically eliminated due to dispersive interactions. The resulting effective Hamiltonian, derived in the appendix, has a similar form to the one of Eq.~(\ref{ham2}) apart from two additional Stark shift terms. By proper choices of parameters can these two terms be made small and the corresponding spectrum then possesses the same properties as Hamiltonian (\ref{ham2}). 

\subsection{Non-zero temperature ground state}\label{ssec3c}

\begin{figure}[ht]
\begin{center}
\includegraphics[width=7cm]{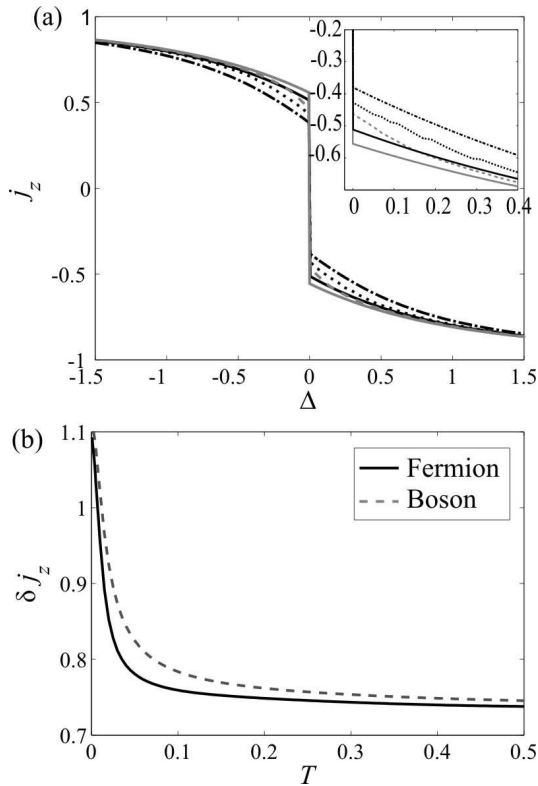}
\caption{The collective atomic inversion $j_z$ (a) for fermions (black lines) and bosons (gray lines) as function of the detuning $\Delta$ and different temperatures, fermions: $T=0$ (solid line), $T=0.02$ (dotted) and $T=0.1$ (dot-dashed) and bosons: $T=0$ (solid line) and $T=0.1$ (dashed line). The inset is a close-up of the curves to the right of $\Delta=0$. The second figure (b) depicts the temperature dependence of the inversion gap $\delta j_z$ of Eq.~\ref{invgap}, for Fermions (solid black line) and bosons (dashed gray line). In both plots, $g=1/2$ and the chemical potential is chosen such that the number of particles is half the lattice number. }
\label{fig5}
\end{center}
\end{figure}  

We turn now to the situation of non-zero temperature and thermal excitations. As argued above, each eigenstate $|\psi_1(q)\rangle$ in the zero temperature ground state undergoes a collective Rabi flip by sweeping the detuning across resonance, which suggests that also for excitations around the Fermi surface or of the Boson ground state will the atomic inversion be discontinuous around $\Delta=0$. Let us introduce the gap function
\begin{equation}\label{invgap}
\delta j_z=\lim_{\Delta\rightarrow0^-}j_z-\lim_{\Delta\rightarrow0^+}j_z
\end{equation}
and study the behavior of $\delta j_z$ for various temperatures $T$. The temperature dependencies of the collective atomic inversion $j_z$ and the inversion gap $\delta j_z$ are presented in Figs.~\ref{fig5} (a) and (b) respectively. The gap is slightly larger for Bosons than for Fermions. Surprisingly, the gap $\delta j_z$ approaches a non-zero value ($\approx0.75$ for this choice of coupling; $g=1/2$) in the large temperature limit. Even at fairly large couplings, this asymptotic value is non-zero. 

\begin{figure}[ht]
\begin{center}
\includegraphics[width=8cm]{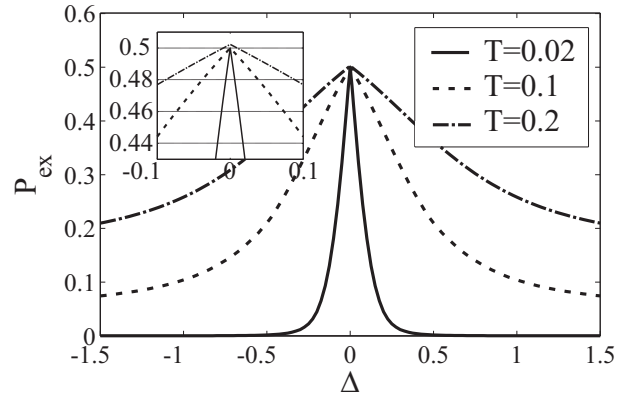}
\caption{Population of excited bands $P_{ex}$, in the case of fermions, as function of the detuning and of temperature; $T=0.02$ (solid line), $T=0.1$ (dashed line) and $T=0.2$ (dot-dashed line). At resonance and moderate temperatures, the excited population approximates 1/2 due to the degeneracy at that point. For large detuning, the gap between the two lowest bands increases causing a decrease in the excited band population. The inset shows a close-up of the population around $\Delta=0$, indicating that for large temperatures $P_{ex}$ exceeds 1/2 at resonance due to occupation of third and higher bands. Here $g=1/2$ and $\xi=1/2$. }
\label{fig6}
\end{center}
\end{figure}   

Another relevant question is the amount of populations of excited quasi momentum states $|\psi_\nu(q)\rangle$ for non-zero temperatures, especially for fermions. The two lowest Bloch bands are identical at resonance, $\Delta=0$, and consequently equally populated. For small temperatures, only states corresponding to these two bands are substantially occupied, while for large temperatures also the third and forth band will be populated. For non-zero detuning, the degeneracy is lifted and the populations of the two lowest bands are no longer balanced. We may define the individual band population 
\begin{equation}
P_\nu=\sum_{q\in(-1,1]}\langle\hat{n}_q^\nu\rangle,
\end{equation}   
where $\langle\hat{n}_q^\nu\rangle=\mathrm{Tr}\left[\rho_T\hat{n}_q^\nu\right]$ is the expectation value of the number operator $\hat{n}_q^\nu$ for the state $\rho_T$ given at temperature $T$. Thus, $P_\nu$ measures the population in band $\nu$, and in particular for $\Delta=0$ do we have $P_i=P_{i+1}$ for $i=1,\,3,\,5,\,...$ due to the degeneracy. The total population of excited bands is given by
\begin{equation}
P_{ex}=\sum_{\nu=2}^\infty P_\nu.
\end{equation}
Note that $P_{ex}$ includes the second band $\nu=2$ which for $\Delta=0$ is degenerate with the lowest band $\nu=1$ and in this special case is the subscript $ex$ (excited) misleading. Naturally, $0\leq P_{ex}<1$ and also $1/2\leq P_{ex}(\Delta=0)<1$. Figure~\ref{fig6} displays $P_{ex}$ as function of $\Delta$ and for three different temperatures. As expected, the excitations increase for a large temperature and decrease for a large detuning. The inset gives the population close to resonance, and it is in particular seen that $P_{ex}(\Delta=0)>1/2$ ($P_{ex}(\Delta=0)=0.5025$ for $T=0.2$) from the fact that the bands $\nu=3,\,4,\,...$ begin to be populated. The shape of $P_{ex}$ seems fairly Lorentzian, but the inset reveals that $P_{ex}$ is indeed not Lorentzian in the vicinity of $\Delta=0$.  Again we restrict the analysis to a filling factor $\xi=1/2$, namely chose the chemical potential $\mu$ such that it results in half filling.

\subsection{Extension to a higher dimensional optical lattice}\label{ssec3d} 
We conclude this section by discussing the situation of a two dimensional optical lattice. The two fields are assumed to share the same wavelength and both interact with the same dipole transition of the atom. The Hamiltonian takes the form
\begin{equation}\label{ham2d}
\begin{array}{l}
\displaystyle{\hat{H}_{2D}=-\frac{\partial^2}{\partial x^2}+}\\ \\
\displaystyle{+\left[\begin{array}{cc}
\displaystyle{\frac{\Delta}{2}} & 2g_x\cos(\hat{x})+2g_y\cos(\hat{y}) \\ 
2g_x\cos(\hat{x})+2g_y\cos(\hat{y}) & \displaystyle{\frac{\Delta}{2}}\end{array}\right].}
\end{array}
\end{equation}
The lowest energy band in the symmetrical case with $g_x=g_y=0.1$ and $\Delta=0.25$ is shown in Fig.~\ref{fig5}. As for the one dimensional case, for $q_x,\,q_y=\pm1/2$ does the lowest energy band attain its maximum. Importantly, multiple minima of the dispersion is also found in the two dimensional case. It is easy to convince oneself that this holds also in three dimensional, and consequently that the topological PT is not limited to the one dimensional lattice.   
  
\begin{figure}[ht]
\begin{center}
\includegraphics[width=8cm]{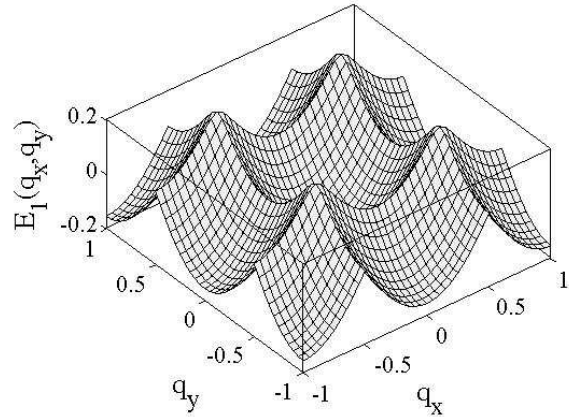}
\caption{First energy band $E_1(q_x,q_y)$ of the two dimensional model (\ref{ham2d}), with equal coupling strengths $g_x=g_y=0.1$ and $\Delta=0.25$. }
\label{fig7}
\end{center}
\end{figure}   

\section{Conclusions}\label{sec4}
In this paper have we studied an ideal gas of ultracold two-level atoms coupled to an optical lattice. It was shown that the coupled two-level character of the problem give rise to novel phenomena, not present in the regular dispersive case of large detuning where one atomic level has been adiabatically eliminated rendering an internal structureless system. In particular, the Brillouin zone is twice as large in this system compared to a internal structureless one and the Bloch bands possess multiple minima. This comes about due to the competition between the various involved terms of the Hamiltonian; kinetic and internal energy. An outcome of the peculiar energy band spectrum is the presence of a topological, or Lifshitz, PT as the detuning is tuned across resonance. This transition is of first order nature and is manifested in the collective atomic inversion. Moreover, it was found to be a non-zero temperature PT. 

To conclude, we showed that quantum PTs can occur in multi-component ultracold atomic systems despite the lack of atom-atom scattering. {\bf It is believed that adding interaction between atoms can give rise to novel phenomena. In such a scenario, two relevant questions directly come to mind: How does the coupled internal structure of atoms affect the phase diagrams of non-coupled models, and how does strong interaction influence the topological PT demonstrated in this paper? The second question was briefly discussed at the end of Sec.~\ref{ssec3b}, and both  questions are currently under investigations for future publications.} We further plan to analyze the present system, and especially the topological PT, in a cavity QED model where the field is treated quantum mechanically.   

\begin{appendix}\label{app}
\section{Effective dissipationless two-level model}
Decay of the excited atomic level will typically set limits on any experimental consideration. One way to minimize such effects is to effectively couple two metastable atomic states using an additional laser. We consider therefore a three-level $\Lambda$-atom with metastable lower states $|1\rangle$ and $|2\rangle$ and excited state $|3\rangle$, with respective energies $E_i$, $i=1,\,2,\,3$. State 1 and 3 are coupled through our optical lattice, while 2 and 3 couple via an external laser whose field amplitude is assumed constant over the atomic sample. The Hamiltonian becomes \cite{adel}
\begin{equation}
\begin{array}{lll}
\hat{H}_\Lambda &  = & \displaystyle{\sum_{i=1}^3E_i\hat{\sigma}_{ii}+\Omega\left(\hat{\sigma}_{23}\mathrm{e}^{i\omega_Lt}+\hat{\sigma}_{32}\mathrm{e}^{-i\omega_Lt}\right)}\\ \\ 
& & \displaystyle{+2g\cos(\hat{x}+\omega_Ot)\left(\hat{\sigma}_{13}+\hat{\sigma}_{31}\right),}
\end{array}
\end{equation}
where $\Omega$ is the external laser coupling amplitude, $\omega_L$ and $\omega_O$ the two field frequencies, and $\hat{\sigma}_{ij}=|i\rangle\langle j|$. The excited state can be adiabatically eliminated if the interaction is dispersive. Thus, we assume a highly detuned configuration, $\Delta_1=E_{33}-E_{22}-\omega_L\gg\Omega$ and $\Delta_2=E_{33}-E_{11}-\omega_O\gg2g$, and at the same time $\Delta_3=|\Delta_1-\Delta_2|$ is of the same order as $\Omega$ and $g$. One then derives, after application of a rotating wave approximation, an effective two-level model for the metastable states, which is given by \cite{adel}
\begin{equation}\label{effham}
\begin{array}{lll}
\hat{H}_{eff} & = & \displaystyle{\frac{\Delta_3}{2}\hat{\sigma}_z-\frac{\Omega^2}{\Delta_1}\hat{\sigma}_{22}+\frac{4g^2}{\Delta_2}\cos^2(\hat{x})\hat{\sigma}_{11}}\\ \\
& & 
\displaystyle{+2g\Omega\cos(\hat{x})\left(\frac{1}{\Delta_1}+\frac{1}{\Delta_2}\right)\hat{\sigma}_x,}
\end{array}
\end{equation}
where $\hat{\sigma}_z=|2\rangle\langle2|-|1\rangle\langle1|$ and $\hat{\sigma}_x=|1\rangle\langle2|+|2\rangle\langle1|$. We have numerically confirmed that the band structure of (\ref{effham}) looks very similar to the ones presented in Fig.~\ref{fig1}. Therefore, our results and conclusions of Sec.~\ref{sec3} are also reproducible for a Hamiltonian such as (\ref{effham}). 

\end{appendix}

\section{Acknowledgements}

We wish to thank Prof. Maciej Lewenstein and Dr. Giovanna Morigi for inspiring discussions.

\end{document}